\documentstyle[12pt]{article}

\title{The end of c=1 CFT}

\author{Kiyohide Nomura	\\
        {\em Department of Physics,} \\
        {\em Tokyo Institute of Technology,} \\
        {\em Tokyo 152, Japan.} \\
}

\date{\today}

\begin{document}

\maketitle

\begin{abstract}
Conformal field theory (CFT) 
with the central charge c=1 is important both in the field theory and
in the condensed matter physics, since it
has the continuous internal symmetry (U(1) or SU(2)) 
and a marginal operator, and it is an effective theory of 
many 1D quantum spin and 1D electron systems.  
So it is valuable to understand how the c=1 CFT models 
become unstable.  
In this paper we discuss an instability of the c=1 CFT,
that is, the transition to the ferromagnetic state.  
For the U(1) CFT case, we find that the spin wave velocity $v$ 
and the critical dimension $K$ 
behave $v, 1/K \propto \sqrt{\Delta_c - \Delta}$
under reasonable assumptions.  
\end{abstract}


\pagebreak

\section{Introduction}

  Conformal field theory (CFT) \cite{B-P-Z} is an efficient theory to
classify universality classes of two-dimensional (2D) classical
systems and 1D quantum systems.  
In particular, the CFT with the central charge c=1 is interesting, since 
it has the continuous internal symmetry and continuously varying exponents 
driven by a marginal operator, and the c=1 CFT is related with 
the Tomonaga-Luttinger liquid \cite{haldane-tl}.  
Therefore it is important to understand how the c=1 CFT models become
unstable.  
Until now, three types of instability for the $c=1$ CFT have been known: 
the usual second order transition around the Gaussian fixed
line \cite{Kadanoff-b}, 
the bifurcation to two c=1/2 CFT lines at the Ashkin-Teller point 
\cite{Kohmoto-dN-K,Yang}, and the
Berezinskii-Kosterlitz-Thouless (BKT) 
transition \cite{Berezinskii,Kosterlitz-T,Kosterlitz,Nomura}.  

In this paper we study another type of the instability of the c=1 CFT,
that is, the transition to the ferromagnetic state.  
For physical systems, we mainly discuss 1D quantum spin systems.  
In the case of strongly correlated 1D electron systems, 
the phase separation at $J_c/t=2.5-3.5$ 
of the 1D $t-J$ model \cite{Ogata-L-S-A} is related with 
the U(1) CFT instability, 
and the transition from the paramagnetic Tomonaga-Luttinger liquid to the
itinerant ferromagnetic phase in the 1D Kondo-lattice model 
\cite{Tsunetsugu-S-U,Ueda-N-T} 
is related with the instability of the SU(2) CFT.  
For the 1D $t-J$ model, numerically it is observed that 
the critical exponent $K_\rho$ for the charge part 
diverges near the phase boundary, 
but the asymptotic behavior of $K_\rho$ is not known.  
For the 1D Kondo-lattice model, there remains a disagreement 
about the mechanism of the transition to the ferromagnetism.  
Using the non-Abelian bosonization, Fujimoto and Kawakami
\cite{Fujimoto-Kawakami} discussed a direct transition from 
the gapless Tomonaga-Luttinger phase to the ferromagnetic phase, 
while White and Affleck \cite{White-Affleck} 
argued that there is an intermediate spin-gap phase.   
We expect this research to contribute a deeper understanding 
for these problems.  

This paper is organized as follows.  In section 2, we discuss an 
instability of the c=1 CFT with the U(1) symmetry, and we examine several
physical examples.  The instability to the
ferromagnetism is well described with the Gaussian model.  
In section 3, we discuss the instability of 
the SU(2) c=1 CFT to the ferromagnetism.  In this case, the mechanism of the
instability is different from the U(1) case, caused by the increase of
the marginally irrelevant coupling.  And the ferromagnetic state
adjacent to the SU(2) c=1 CFT phase has two soft modes 
at $k=0$ and $k=\pi$.  Finally, a summary is given in section 4.

\section{U(1) symmetry case}

  The ${\rm U(1)\times Z_2 \times Z_2}$ symmetric CFT model is
described generally with a sine-Gordon Lagrangian \cite{Nomura}: 
\begin{equation}
	{\cal L}={1 \over 2\pi K} (\nabla \phi)^2 
	+{y_\phi \over 2\pi \alpha^2 } \cos \sqrt{2} \phi,
	\label{eqn:SGlag}
\end{equation}
where we make the identification 
$\phi \equiv \phi +2 \pi/\sqrt{2}$.
The dual field $\theta$ defined as
\begin{equation}
	\partial_x \phi = -\partial_y(i K \theta), \:
	\partial_y \phi = \partial_x(i K \theta), 
\end{equation}
has the internal U(1) symmetry with the identification 
$\theta \equiv \theta + 2 \pi/\sqrt{2}$.  
Besides the U(1) symmetry, the sine-Gordon model has discrete symmetries
under the transformations 
$(z, \phi, \theta) \rightarrow (z, -\phi, -\theta)$ and 
$(z, \phi, \theta) \rightarrow (\bar{z}, \phi, -\theta)$.  
So that for $y_\phi=0$, it has 
the ${\rm U(1) \times U(1) \times Z_2 \times Z_2}$ symmetry, 
and for $y_\phi \neq 0$, the ${\rm U(1) \times Z_2 \times Z_2}$ symmetry.  

First we consider the Gaussian model $y_\phi=0$ \cite{Kadanoff-b}.  
The self-dual point 
$K=1$ corresponds to the SU(2) $k=1$ Wess-Zumino-Witten (WZW) model, 
$K=4$ corresponds to the BKT multicritical point.  With a finite $y_\phi$
and $K<4$, there is a mass generation with the second order transition,
whereas $K>4$, there is an extended massless region between 
the two BKT lines($y_\phi=\pm y_0, y_0 \equiv 2(K/4-1) $), 
where all the points are renormalized to the Gaussian fixed line.  
In the limit of $K \rightarrow + \infty$, there appears another type 
of instability.  
The critical dimensions for the operators
\begin{equation}
	O_{n,m} = \exp (i n \sqrt{2} \phi) \exp (i m \sqrt{2} \theta)
\end{equation}
are
\begin{equation}
	x_{n,m} = \frac{1}{2} \left( n^2 K + \frac{m^2}{K} \right),
	\label{eq:gauss-critical}
\end{equation}
so that $x_{0,m} \rightarrow 0$ for $K \rightarrow \infty$, 
which implies a ferromagnetic long-range order.  

To clarify the situation, we consider the finite lattice spin system.  
For a finite $L$ system with periodic boundary conditions, 
the excitation energies are related to 
the critical dimensions \cite{Cardy}
\begin{equation}
	\Delta E_{n,m} (L) = \frac{2\pi v}{L} x_{n,m},
\end{equation}
and the correction to the ground state energy is described 
by the conformal charge $c$ 
\cite{blote-cardy,affleck}
\begin{equation}
	E_g (L) = e_0 L - \frac{\pi v}{6 L} c,
\end{equation}
where $v$ is the spin wave velocity.  
Since the operator $O_{0,m}$ takes $|m| \leq S$ on one site, 
for the finite $L$ chain with spin $S$ 
the quantum number $m$ takes integers $|m| \leq SL $.  
In the $K \rightarrow \infty $ limit, the excitation energies 
$\Delta E_{0,m}$ become degenerate with the ground state energy, 
consequently there is a $2SL + 1$ degenerate ground state, 
which suggests an SU(2) ferromagnetic ground state.  
On the other hand, the ground state in the $K<0$ region is 
a two-fold degenerate $S^z_T \equiv \sum S^z_j = \pm SL$ ferromagnetic state. 

In general, excitation energies are limited by the band-width
($\propto L$), 
and $\Delta E_{n,0} = \pi n^2 v K /L (n_{max} \propto L)$, 
so it should be $v \propto 1/K$. 
This explains the change of the dispersion curve from 
the type $\omega \approx v |q|$ in the XY region 
to the $\omega \propto q^2$ on the SU(2) ferromagnet, 
and that the ground state energy of the ferromagnet does not
depend on the system size.  

It is possible to consider the confluent (or rather effluent) point of 
the BKT transition line and the SU(2) ferromagnetic line.  
This is a divergent point of all the critical lines which renormalize to 
the Gaussian fixed points $4<K<\infty$, 
and this is an infrared unstable fixed point.  
In order to explain this feature, 
at this effluent point 
the dispersion curve should be flat and there should be a highly
degenerate ground state $\propto \exp (c L)$ \cite{berker-kadanoff}.

\subsection{Physical models}

\subsubsection{bond-alternating XXZ spin chain}

For a physical model, we treat the bond-alternating XXZ spin chain
\begin{equation}
	H= - \sum(1+\delta (-1)^j)( S^x_j S^x_{j+1}+S^y_j S^y_{j+1}
		+\Delta S^z_j S^z_{j+1}).
	\label{bond-alterXXZ}
\end{equation}
This model has a massless XY phase close to the $\Delta \leq 1$, 
and becomes massive after the BKT transitions (see Fig.1) 
\cite{Kohmoto-dN-K,Kitazawa-N-O}.  
When $\Delta =1, |\delta|<1 $, this model has an SU(2) ferromagnetic 
long-range order with a  $2SL +1$ degenerate ground state, 
while for $\Delta >1, |\delta|<1$, 
it has a two-fold degenerate ferromagnetic ground state
($S^z_T = \pm SL$).

At the points $\Delta=1, \delta= \pm 1$, the model (\ref{bond-alterXXZ}) is
considered as the product of SU(2) ferromagnetic pairs, so it has 
a $(4S + 1)^{L/2}$ degenerate ground state.  
This is a locally degenerate state, and the dispersion curve is flat, 
consequently the points $\Delta =1, \delta = \pm 1$ 
are considered as the effluent point.  
On the lines $\Delta>1, \delta= \pm 1$, the model (\ref{bond-alterXXZ}) is
considered as the product of $Z_2$ ferromagnetic pairs, so it has 
a $2^{L/2}$ degenerate ground state.  
This line separates the ferromagnetic phase from a type of N\'eel 
phase (up-up-down-down) \cite{Yamanaka-H-K}.  

In the case of $S=1/2,\delta=0$, exact results are known 
from the Bethe Ansatz \cite{cloizaux,yang-yang,luther-peschel}.  
The critical dimension is a function of the anisotropy $\Delta$
\begin{equation}
	1/K = \eta_{xy} = 1/ \eta_z = (1/\pi) \arccos (\Delta) 
	\approx (1/\pi) \sqrt{2(1-\Delta)},
\end{equation}
and the spin wave velocity is 
\begin{equation}
	v = (\pi/2) \sin (\arccos (\Delta))/(\pi- \arccos \Delta)
	\approx (1/2) \sqrt{2(1-\Delta)}.
\end{equation}
Therefore, near $\Delta=1$ the spin wave velocity behaves 
$
	v \propto 1/K
$, 
as expected.  
Note that in the XY phase 
the fully ferromagnetic state has the excited energy near 
$\Delta=1$ 
\begin{equation}
	\Delta E_{0,\pm L/2} = \frac{\pi v}{L}\frac{(L/2)^2}{K} 
	= \frac{L}{4}(1-\Delta),
	\label{eq:1st-o}
\end{equation}
which is consistent with the fact that the transition between the XY 
and the ferromagnetic phase is the first order transition.  
The asymptotic behavior (\ref{eq:1st-o}) applies not only to 
the integrable line but also to the whole 
$
|\delta|<1, \Delta \approx 1
$ 
region (see appendix).  

For the general spin $S$ case, in the XY phase 
the fully ferromagnetic states ($S^z_T = \pm SL$) have the excited energy 
\begin{equation}
	\Delta E_{0,\pm SL} = S^2 L (1-\Delta)[1 +O(1-\Delta,1/L)],
	\label{eq:1st-o2}
\end{equation}
(see appendix).  Comparing this with the Gaussian model arguments
(\ref{eq:gauss-critical}), we obtain 
\begin{equation}
	\Delta E_{0,\pm m} = \frac{m^2}{L} (1-\Delta) 
	= \frac{\pi v}{L}\frac{m^2}{K}.
	\label{eq:1st-o3}
\end{equation}
Therefore, assuming $v \propto 1/K$ near $\Delta=1$, 
we obtain $v, 1/K \propto \sqrt{1-\Delta}$.  
And the energy gap between the fully ferromagnetic state and 
the one-spin flip state ($S^z_T = \pm (SL-1)$) is 
\begin{equation}
	\Delta E_{0,\pm SL}-\Delta E_{0,\pm (SL-1)}=
	\frac{2SL-1}{L}(1-\Delta) \approx 2S (1-\Delta),
\end{equation}
which is consistent with the simple spin wave calculation.  

In the neighborhood of the effluent points ($\Delta=1, |\delta|=1$), 
eq. (\ref{eq:1st-o3}) implies that the spin-wave velocity 
on the BKT line behaves as 
$v = (4/\pi) (1-\Delta)$, since the coupling $K$ renormalizes to $4$ 
on the BKT line.  
Numerically it is observed that the BKT lines are 
$1-\Delta \propto 1-|\delta|$ close to the effluent points.  
Combining this with Eq. (\ref{eq:1st-o3}), we obtain
\begin{equation}
	\frac{1}{K} \approx \frac{1}{\pi f(S)}\sqrt {\frac{1-\Delta}{1-|\delta|}},
\end{equation}
and
\begin{equation}
	v \approx f(S) \sqrt {(1-\Delta) (1-|\delta|)}.
\end{equation}
Since the band-width of the model (\ref{bond-alterXXZ}) 
is proportional to $S^2$, the asymptotic behavior of $f(S)$ 
is considered as $f(S) \propto S$.

\subsubsection{XXZ spin chain with a staggered magnetic field}

Next we consider the XXZ spin chain with a staggered magnetic field 
\begin{equation}
	H= - \sum ( S^x_j S^x_{j+1}+S^y_j S^y_{j+1} + \Delta S^z_j S^z_{j+1} 
		+ h_s (-1)^j S^z_j).
	\label{stagger}
\end{equation}
For the S=1/2 case, this model was studied by Alcaraz and 
Malvezzi \cite{Alcaraz-Mal}.
The phase diagram is similar to the bond-alternating XXZ spin chain.  
The boundary between the massless XY phase and the ferromagnetic phase 
is $\Delta_c = \sqrt{h_s^2 +1}$.  Although on this line Hamiltonian
(\ref{stagger}) is not SU(2) invariant, 
all the lowest states in the sectors with 
$S^z_T = - L/2, -L/2+1, \cdots, L/2$ are degenerate \cite{Alcaraz-Mal},
which is consistent with the expectation based on the sine-Gordon 
model that at the boundary between the massless XY
phase and the ferromagnetic phase, the degeneracy of the ground
state is $2SL + 1$.  

About the asymptotic behavior of $K,v$ near $\Delta_c$, 
using the ground state wavefunctions at $\Delta_c$ \cite{Alcaraz-Mal} 
and the similar method as the appendix, we can show that 
the fully ferromagnetic states 
($S^z_T = \pm L/2$) have the excited energy
\begin{equation}
	\Delta E_{0,\pm L/2} = \frac{\pi v}{L}\frac{(L/2)^2}{K} 
	= \frac{L}{4}(\Delta_c-\Delta).
\end{equation}
Therefore, we obtain $v,1/K \propto \sqrt{\Delta_c - \Delta}$

On the other hand, there is not the point with the highly degenerate
ground state $\approx \exp(cL)$, so that the BKT line and the
SU(2)-like ferromagnetic transition line will not intersect.  

For the general spin $S$ case, the XY-ferromagnetic boundary is estimated
as $\Delta_c=\sqrt{1+h_s^2/4S^2}$, based on the stability of the
ferromagnetic state to the one spin-flip state.  
It is possible to show that the degeneracy on this boundary is 
$2SL+1$ \cite{Alcaraz-priv}.

\subsubsection{XXZ spin chain with a single-ion anisotropy}

In connection with the Haldane conjecture \cite{Haldane}, 
the XXZ spin chain with a single-ion anisotropy (spin $S\geq 1$)
\begin{equation}
	H= \sum S^x_j S^x_{j+1} + S^x_j S^x_{j+1} +\Delta S^x_j S^x_{j+1} 
		+ D ( S^z_j)^2.
	\label{singleION}
\end{equation}
was studied extensively by bosonization \cite{Schulz}, 
by numerical calculations for the S=1 \cite{Botet,Glaus} case, 
and for the S=2 case \cite{Jolicoeur}.  
In the large $D\rightarrow +\infty$ limit, 
the N\'eel phase($D \gg \Delta$), the large-D phase, the XY phase and
the ferromagnetic phase($D \ll -\Delta$) are considered.  
Using bosonization, Schulz discussed that between the large-D phase 
and the Haldane phase there is always the intermediate massless XY phase, 
whereas from the numerical calculations it seems there is a direct first-order
transition from the large-D to the ferromagnetic phase.  
However, since the BKT transition line can be determined by the crossing of
the $S^z_T=0$ and the $S^z_T=\mp 4$ excitations for finite systems 
\cite{Nomura}, 
and the ferromagnetic boundary is determined by the crossing of 
the lowest $S^z_T=0$ state and the $S^z_T = \pm SL$ state, 
so that these two lines cannot intersect with each other.  

On the ferro-XY boundary, note that the two spin-flip bound state
makes the ferromagnetic state unstable prior to the one-spin 
flip state, because of the single-ion anisotropy term
\cite{Papanicolau}.

\section{SU(2) symmetry case}

The $c=1$ conformal field model with the SU(2) symmetry is described 
by the SU(2) $k=1$ Wess-Zumino-Witten (WZW) model 
\cite{Knizhnik,Affleck-g-s-z}.  
In order to inhibit the SU(2) symmetric relevant interaction from producing
a mass, an additional $Z_2$ symmetry, the symmetry of translation by one
site, is needed \cite{Affleck85}.  In this case the 
relevant field corresponds to the state with momentum $\pi$, so that 
the translationally invariant perturbation does not produce a mass.  
In addition to the relevant operator, there is a marginal operator 
$\mbox{\boldmath$J$}_L \cdot \mbox{\boldmath$J$}_R$.  
This will be marginally relevant or irrelevant
depending on the sign of the coupling constant.  

Perturbating the WZW model with the marginally irrelevant term, 
we obtain the correction to the eigenvalue for the finite 
system $L$ with periodic boundary conditions \cite{Cardy86,Cardy86b}
\begin{equation}
	\Delta E_n (L) = \frac{ 2 \pi v}{L} 
	\left( x_n + \frac{2 b_n}{b} \pi b g \right),
\end{equation}
where $x_n$ is the critical dimension, $v$ is the spin wave velocity, 
$b_n$, $b$ are related to the operator product expansion
coefficients.  
The marginal coupling $g$ is renormalized as
\begin{equation}
	\frac{d (\pi b g)}{d l} = - (\pi b g) ^2 
	+ O (g^3),
\end{equation}
where $l \equiv \ln L$.  In the lowest order, the effective coupling
constant $g(L)$ is renormalized as 
\begin{equation}
	\pi b g (L)= \frac{\pi b g_0}{1 + \pi b g_0 \log L},
\end{equation}
where $g_0$ is a system dependent renormalization constant.  
In general, however, there are the corrections such as 
$\ln (\ln L)/(\ln L)^2$ from higher order terms.  

In the space with the total spin $\mbox{\boldmath$S$}^2 = n(n+1)$ 
and the wavenumber 0 for $n$ even, $\pi$ for $n$ odd, 
the lowest excitations are characterized with 
$ x_n = n^2/2, b_n/b = -n^2/8$ \cite{Affleck-g-s-z}, that is,
\begin{equation}
	\Delta E'_n (L) = \frac{2 \pi v}{L}\frac{n^2}{2}
	\left( 1-\frac{1}{2}\pi b g \right),
	\label{su2ex}
\end{equation}
When $g_0 \rightarrow 0+$, 
the logarithmic corrections decrease, 
and the system renormalizes to the free $k=1$ ${\rm SU(2)\times SU(2)}$ 
WZW model.  When $g_0<0$, the coupling becomes marginally relevant, 
which causes the $Z_2$ symmetry breaking.  

On the other hand, another type of the instability may occur with the
increase of $g_0$.  
When $\pi b g = 2$, the excitations (\ref{su2ex})
become 0, and these excitations (wavenumber $0$ and $\pi$) become 
degenerate with the ground state.  The degeneracy is estimated to be 
\begin{equation}
	\sum_{n=0}^{SL} (2n+1) = (SL +1)^2.
\end{equation}  
This can be interpreted as the boundary
with the ferromagnetic phase, which has soft modes at $k=0,\pi$ and the
dispersion curve is characterized by $\omega \propto k^2, (k-\pi)^2$.  
However, there is a problem in the above picture, since the coupling
$g$ is size dependent.  One solution for this problem is that 
$\pi b g^* =2$ is the unstable fixed point $d  g^* / dl =0$.

\subsection{Heisenberg chain with next-nearest-neighbor interactions}

For a physical model, we consider the Heisenberg chain with 
next-nearest-neighbor interactions
\begin{equation}
	H=\sum ( \cos \theta \mbox{\boldmath$S$}_j \cdot 
		\mbox{\boldmath$S$}_{j+1}
		+ \sin \theta  \mbox{\boldmath$S$}_j \cdot 
		\mbox{\boldmath$S$}_{j+2}).
\end{equation}
The stability of the SU(2) ferromagnetic state can be considered 
as follows.  
The one spin-flip $S^z_T = SL -1$ excitation spectrum from the fully
ferromagnetic state $S^z_T = SL$ is 
\begin{equation}
	\omega(k) = 2S ( \cos \theta (\cos k -1) 
	+ \sin \theta (\cos 2k -1)).
\end{equation}
This spectrum shows the instability in two ways.  First 
at $\theta=-\pi/2$, the spectrum $\omega(k) = 2 S ( 1 - \cos 2 k)$ 
has two minima at $k=0$ and $k=\pi$.  
At this point the ground state is $(SL +1)^2$-fold degenerate.  
When $\theta > -\pi/2$, $\omega(\pi) = -4 S \cos \theta$ becomes negative, 
consequently the ferromagnetic state becomes unstable.  
The instability at this point is explained with the mechanism in the 
previous subsection.  
On the other hand, at $\theta_{FF}= \arctan ( -1/4)$, the curvature 
of the spectrum near $k=0$ changes from positive to negative, 
so $\theta<\theta_{FF}$, the ferromagnetic state becomes unstable.  
Bader and Schilling \cite{Bader-Schilling} proved that in the region 
$\theta_{FF}< \theta <-\pi/2$ the ground state is ferromagnetic, and
outside this region, the ferromagnetic state becomes unstable.  

For the spin $S=1/2$, the phase diagram of this model is considered 
as follows (see Fig. 2).  
The instability at $\theta_{FF}$ was first pointed out by
\cite{Ono}, and the degeneracy at this point was discussed by 
\cite{Bader-Schilling,H-N-Natsume}.  
Between $\theta_{FF}<\theta< -\pi/2$, there is an SU(2) ferromagnetic
ordered phase.  
Above $\theta > \theta_c= \arctan (0.2411)$ \cite{oka-nom,nom-oka}, 
there is a dimer-ordered phase, which is
characterized by the two-fold degenerate ground state and 
the dimer long-range order.  
At $\theta_{MG} = \arctan (1/2)$, the exact dimer ground state was
obtained \cite{majumdar-ghosh}.  

In the region $\theta_{MG} < \theta < \theta_{FF}$, relatively few things
are known \cite{Tonegawa-Harada,Bursil-Gehring,White-Affleck}.  
Between $\theta_{MG} < \theta < \pi/2$, although there remains the
dimer long range order and the system has an energy gap, the spin
correlation function has an incommensurate pitch angle 
$\pi/2 < \phi <\pi$.  
Between $\pi/2 < \theta < \theta_{FF}$, it is believed that the system
is gapless and it has 
an incommensurate pitch angle $0 < \phi <\pi/2$ \cite{White-Affleck}.  

Between $-\pi/2< \theta <\theta_c$, there is an SU(2) spin-fluid phase,
characterized by the massless excitation and the power law decay of 
the correlation functions, and the universality class of this phase is 
of the SU(2) $k=1$ WZW type.  
The critical point $\theta_c$ belongs to the 
${\rm SU(2) \times SU(2)}$ $k=1$ WZW universality class, where the
coupling constant for the marginal field changes the sign, and all the
points in the SU(2) spin-fluid region renormalize to this point $\theta_c$.  

For the general spin case, from the Lieb-Schultz-Mattis theorem 
\cite{Lieb-S-M} and its extension to the general spin \cite{Affleck-Lieb},  
in the region $-\pi/2< \theta \leq 0$, it is proved that the ground
state is a unique singlet; moreover for the $S$ half-odd integer case, 
the excitation gap decreases $O(1/L)$ with increasing system size $L$.  
Note that the second theorem applies in the case when the ground state
is a unique singlet, and that the constructed excited state has the $-1$ 
quantum number under the discrete $S^z_i \rightarrow -S^z_{-i}$ 
transformation.  
Therefore, for the infinite limit, two possibilities can be considered.  
One is that the excitation spectrum is continuous.   
Another possible situation is that 
the spontaneous symmetry breaking occurs under the 
$S^z_i \rightarrow -S^z_{-i}$ transformation and the ground state
becomes 2-fold degenerate with an energy gap \cite{Affleck-les-houches}.  
The latter case corresponds to the N\'eel state or the dimerized state.  
For the isotropic case, only the dimerized state is possible.  
Therefore, for the general half odd integer spin case, 
we expect the similar phase diagram to the S=1/2 case, 
consisting of the ferromagnetic region, the SU(2) spin-fluid region, 
and the dimer region.

\section{Summary and discussions}

We have studied the instability of the c=1 CFT 
for the U(1) symmetry case and the SU(2) symmetry case.  

In the case of the U(1) symmetry, considering that the excitation
energies are limited by the band-width, 
we showed that the spin wave velocity is inverse to the
Gaussian coupling $v \propto 1/K$.  In addition, since the transition
from the XY phase to the ferromagnetic phase is the first order
transition, the spin wave velocity and the Gaussian coupling behaves 
as $v \propto \sqrt{1-\Delta}, K \propto 1/\sqrt{1-\Delta}$.  

It is possible to consider the effluent point of the BKT transition
and the ferromagnetic region, or the disappearance of the intermediate
massless XY phase.  At this point the dispersion curve should be flat, 
which suggests a locally degenerate ground state.  
In the case of the bond-alternating XXZ chain, 
$\Delta=1, \delta = \pm 1$ are the effluent points where
the XY phase disappears.  For other models, such as the XXZ spin
chain with staggered fields and the XXZ chain with the single-ion anisotropy, 
there is no such a point with a flat dispersion, therefore there is
always the intermediate XY phase between the ferromagnetic phase and
the massive singlet phase.  

About the 1D $t-J$ model, since the low-energy behavior of the charge
part is described by the U(1) c=1 CFT, and the critical exponent 
$K_\rho$ diverges near the phase separation, 
we expect that the asymptotic behavior of $1/K_\rho$ and $v_c$ is
proportional to $\sqrt{J_c - J}$.  
As for the effluent point, all the critical lines 
$1/2 < K_\rho < \infty$ converge to the point 
$ J/t =2, n=0$, therefore this point is considered to be a type of 
the effluent point.  However, this is a different type from the
effluent point discussed in section 2.  
Since the low density limit in the 1D $t-J$ model
corresponds to the neighborhood of the saturation magnetization 
in the 1D XXZ spin model, the behavior close to 
the point $J/t=2,n=0$ belongs to 
the same type as the point $\Delta=1, S^z_T=\pm SL$ 
\cite{Haldane80}.

In the case of the SU(2) symmetry, we discussed that 
the increase of the marginal irrelevant coupling 
causes the instability from the SU(2) $k=1$
Wess-Zumino-Witten model to the SU(2) symmetric ferromagnetic state.  
From the other side, the SU(2) ferromagnetic state becomes unstable, 
since in addition to the usual ferromagnetic excitation 
$\omega \propto k^2$, 
there appears another soft mode at $k=\pi$: $\omega \propto (k-\pi)^2$.   

For the 1D Kondo lattice model, considering the excitation spectrum 
in the ferromagnetic region, we will determine whether there is the
intermediate spin gap phase or not between the itinerant ferromagnetic
phase and paramagnetic Tomonaga-Luttinger phase.

\section{Acknowledgements}

I am grateful to Francisco Alcaraz, Norio Kawakami, Masao Ogata, 
and Kazuo Ueda for fruitful discussions.  
I would like to thank Atsuhiro Kitazawa to show me numerical data 
prior to publication.

\appendix

\section{Appendix}

Here we show that the relation (\ref{eq:1st-o}) applies not only to the
integrable case $S=1/2,\delta=0$ but also to the non-integrable case
$S$ arbitrary, $|\delta|<1$.  

Close to $\Delta=1$, we can estimate the ground state energy 
by the perturbation.  
We treat the SU(2) ferromagnetic term 
\begin{equation}
	H_0=-\sum (1+ (-1)^j \delta) \mbox{\boldmath$S$}_j \cdot
	\mbox{\boldmath$S$}_{j+1},
\end{equation}
as a free part, and the anisotropic part
\begin{equation}
	H_1=-(\Delta-1) \sum (1+ (-1)^j \delta) S^z_j S^z_{j+1},
\end{equation}
as a perturbation.  

The energy of the fully ferromagnetic state $S^z_T= \pm SL $ is exactly 
\begin{equation}
	E_{ferro} = - \Delta  S^2 L.
\end{equation}
For $\Delta <1$, the ground state has the quantum number $S^z_T=0,q=0$.  
At $\Delta=1$, the ground state wave function in the $S^z_T=0,q=0$ 
space is derived from the fully ferromagnetic state
\begin{equation}
	| \phi \rangle = (S^-_T)^{SL}| S^z_T=SL\rangle,
\end{equation}
where $S^\pm_T \equiv \sum S^{\pm}_j$.  
The zero-th order energy is given by 
\begin{equation}
	H_0 |\phi \rangle = -S^2 L | \phi \rangle.
\end{equation}
To calculate the first order perturbation, 
we have to evaluate 
$\langle\phi| S^z_j S^z_{j+1} |\phi\rangle$.  
Since $| \phi \rangle$ is invariant under the permutation of 
the lattice sites $\{ j \}$, we obtain
\begin{equation}
	\langle \phi | S^z_i S^z_j | \phi\rangle = const. 
	\qquad
	\mbox{for any $i \neq j$},
\end{equation}
therefore, 
\begin{eqnarray}
	\langle \phi | ( \sum S^z_i )^2 | \phi\rangle &=& 0
	\nonumber	\\
	&=& \sum_{i\neq j} \langle \phi | S^z_i S^z_j | \phi\rangle 
	+ \sum_i \langle \phi | (S^z_i )^2 | \phi\rangle
	\nonumber	\\
	&=& L(L-1)  \langle \phi | S^z_j S^z_{j+1} | \phi\rangle 
	+ L \langle \phi | (S^z_i)^2 | \phi\rangle.  
\end{eqnarray}
Considering 
$
\langle \phi |(S^z_i)^2|\phi\rangle /\langle \phi|\phi \rangle \propto
S^2
$,
we obtain 
$
\langle\phi| S^z_j S^z_{j+1} |\phi\rangle
/\langle \phi|\phi \rangle \propto -S^2/L
$.  
Then, the first order perturbation is 
\begin{eqnarray}
	E_1 &=& \langle \phi | H_1 | 
	\phi\rangle /\langle \phi | \phi \rangle	\nonumber	\\
	&=& -(\Delta-1)
	\sum (1+ (-1)^j \delta) \langle \phi| S^z_j S^z_{j+1} | \phi \rangle
	/\langle \phi| \phi \rangle
	\nonumber	\\
	&\propto& -(\Delta -1) S^2.
\end{eqnarray}
Therefore, the energy gap between the singlet ground state and the
fully ferromagnetic state is
\begin{equation}
	\Delta E_{0,\pm SL} = S^2 L (1-\Delta) 
	[1 +  O(1-\Delta,1/L)].
\end{equation}

\pagebreak

\end{document}